\documentclass[runningheads]{llncs}
\usepackage{graphicx}
\begin{document}
\title{Towards A Double-Edged Sword: Modelling the Impact in Agile Software Development}
\titlerunning{ Modelling the Impact in Agile Software Development}
%
\author{Michael Neumann\inst{1}\orcidID{0000-0002-4220-9641} \and
\\ Philipp Diebold\inst{2,3}\orcidID{0000-0002-3910-7898}}
\authorrunning{M. Neumann and P. Diebold}
%
\institute{University of Applied Sciences Hannover, Ricklinger Stadtweg 120, 30459 Hannover, Germany 
\email{michael.neumann@hs-hannover.de}\\
\and
Bagilstein GmbH,  Im Niedergarten 10,  55124 Mainz , Germany
\email{philipp.diebold@bagilstein.de}\\
\and
IU Unternational University, Juri-Gagarin-Ring 152, 99084 Erfurt, Germany\\
\email{philipp.diebold@iu.org}}
\maketitle              
\begin{abstract}
Agile methods are state of the art in software development. Companies worldwide apply agile to counter the dynamics of the markets.  
We know, that various factors like culture influence the successfully application of agile methods in practice and the sucess is differing from company to company.  
To counter these problems, we combine two causal models presented in literature: The Agile Practices Impact Model and the Model of Cultural Impact.
In this paper, we want to better understand the two facets of factors in agile: Those influencing their application and those impacting the results when applying them. 
This papers core contribution is the Agile Influence and Imact Model, describing the factors influencing agile elements and the impact on specific characteristics in a systematic manner. 

\end{abstract}
\section{Introduction}
In the last decades, agile software development (ASD) have gained a lot of research interest (e.g., ~\cite{Diebold.2014,Schon.2017}. Today, agile methods are used in a variety of contexts (organization, industry, region, ...) with different motivations~\cite{VersionOne.2023}. Organizations want to improve product quality, increase the speed of delivery of product increments, or improve predictability. It is therefore not surprising that the question of how to successfully apply agile methods has been investigated~\cite{Chow.2008}, which have led to an understanding of success factors. On this basis, (causal) models have been defined to systematically describe the influences in the planned or existing application of agile methods (e.g., \cite{Diebold.2015a,Neumann.2023}). Two perspectives may be distinguished here: a) The influence on agile practices in relation to their successful application. b) The effects of the application of agile practices on product or project characteristics. Below, we will focus on two specific models considering these two perspectives.

Today, we know that social facets are important for the success when using agile methods as these facets guide the behaviour of people, e.g., how they communicate and act~\cite{Smite.2020}. Also, aspects wrt.\ agile culture are relevant for the successfull use of agile methods~\cite{Kuchel.2023}. 

To be more precise, specific models were presented explaining the influences of social facets like cultural characteristics on agile methods in a systematic manner in the past.
One is the Model of Cultural Impact on Agile Methods (MoCA)~\cite{Neumann.2023}. It describes cultural influences on the use of agile methods on a systematic basis.
Another model, considering primarly the second perspective is the Agile Practices Impact Model~\cite{Diebold.2015a} aiming to provide a systematic description of the impact of agile practices on specific process improvement goals like e.g. (product) quality, development costs, or time.

However, for the current understanding of the influences and impacts on agile methods the available models, underlying theories and empirical findings do not cover or combine both perspectives. Nevertheless, this knowledge is of high importance as we see the need for a bigger picture supporting researchers and practictioners to find well-suited agile practices for their context considering their specific needs.

This motivated us in a first step to combine our models in order to cover the both mentioned perspectives. This paper presents the Agile Influence and Impact Model (AIIM) aiming to provide a solution for the explained challenges.

This paper is structured as follows: In Section~\ref{APIM}, we give a brief introduction of the APIM, followed by a description of MoCA. The core contribution of this paper is the Agile Influence and Impact Model, which we introduce in Section~\ref{AFM}. Finally, the paper closes with a conclusion in Section~\ref{Conclusion}.

\section{Related work}
\label{APIM}
\subsection{Agile Practices Impact Model} The APIM model was created as a basis for an agile capability analysis. The model described the impact of agile practices on the specific impact characteristics, which are detailed as process improvement goals. Even if the model considers the impact on agile practices, it focuses more on the outcome perspective. Thus, the scientific ground for the APIM are agile practices, the impact characteristics and the impact association between them, which is more specified using Influence Factors. The impact between both aspects is defined as binary in terms of a positive or negative impact. 

\subsection{Model of Cultural Impact on Agile Methods} 
\label{MoCA}
The MoCA model was defined to provide a systematic description of cultural influences on agile practices. The scientific ground for MoCA are the Cultural and Agile Elements dimensions and the specified impact between them. The cultural dimension consists of specific characteristics based on often used cultural models in Software Engineering. The agile elements dimension was created using the results from a tertiary study aiming to provide an up-to-date list of agile practices~\cite{Neumann.2022}. The influence between both dimensions is described as positive or negative, in terms of the application of the agile element with regard to the guideline in which it is defined. 

\section{The Agile Influence And Impact Model}
\label{AFM}
\begin{figure*}[htb]
\centering
\includegraphics[scale=0.66]{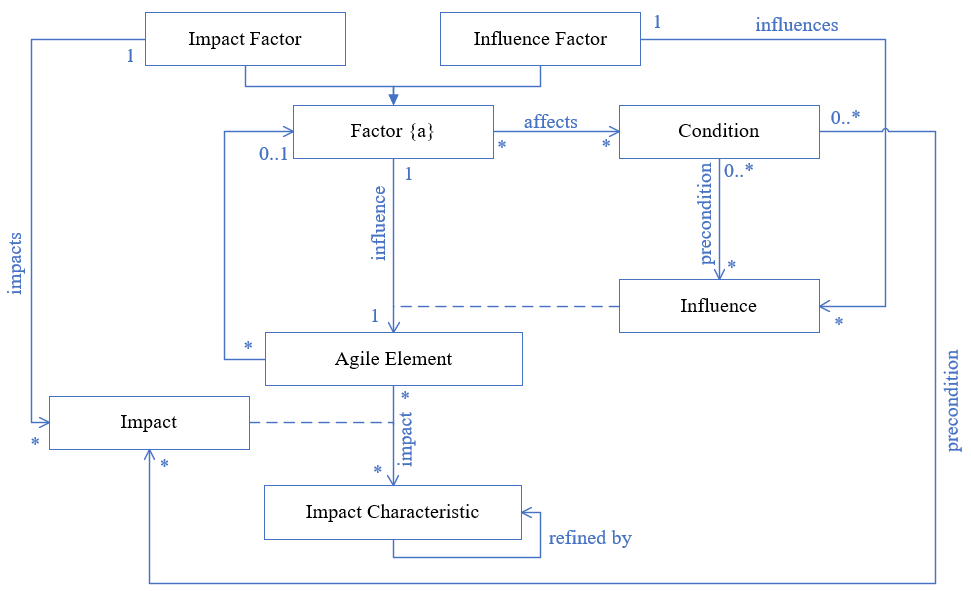}
\caption{Agile Influence and Impact Meta-Model}\vspace*{-1em}\label{fig1:AFM_Meta}
\end{figure*}

In this Section, we provide an explanation of the formal structure on a presented meta-model in Figure~\ref{fig1:AFM_Meta}.

Similar to the two underlying models, we decide to use UML as it fits to the needs explaining the formal structure of the AIIM. Below, we explain the classes and the relationships between them.

An \textit{Agile Element}\footnote{using the MoCA wider definition of the elements of agile methods proposed in~\cite{Neumann.2021a}} is an abstract description of agile activities, roles, and artifacts regardless of their relationship to the different agile methods guidelines. Also, Agile Activities are abstract agile practices as defined in the paper by Diebold and Zehler~\cite{Diebold.2015a}. 

\textit{Factor:} A Factor can be a specific \textit{Influence Factor} or \textit{Impact Factor}. We decided to use a generalized structure as we wanted clearly differ between Impact Factors and Influence Factors to be able to consider both perspectives of influences on agile practices and impacts of agile practices. A \textit{Condition} may applies as a precondition for a specific \textit{influence} from a \textit{Factor} on one \textit{Agile Element}. This influence is not binary in terms of positive or negative (similar to the impact, which we understand as not binary). We assume that an influence of one Influence Factor (e.g., a cultural value) on an \textit{Agile Element} is defined based on the expected application of the \textit{Agile Element} with respective to the guideline in which this practice is defined. An \textit{Impact} is represented by an Impact Factor on an Impact Characteristic, which are often related to process improvement goals, like Development cost or time\cite{Diebold.2015a}.

\section{Conclusion and Future Work}
\label{Conclusion}
In this paper, we present the Agile Factor Model: A meta-model describing both perspectives of influences and impacts with regard to agile methods. The model was created based on the combination of the existing models. 

In a next step, we aim to define how the model can be applied using examples from our previous models. The vision of the AIIM is to provide a theory which can be applied in real-world settings. Thus, we aim in further step to define an application process for the AIIM to be able to evaluate the new model in real-world settings. 

%
%
%
 \bibliographystyle{splncs04}
 \bibliography{references}

\end{document}